\documentclass{article}
\usepackage{spconf,amsmath,graphicx}

\usepackage{amssymb}
\usepackage{type1cm}
\usepackage{multirow}

\title{Vectorization of hypotheses and speech for faster beam search \\ in encoder decoder-based speech recognition}

\name{Author(s) Name(s)\thanks{Thanks to XYZ agency for funding.}}
\address{Author Affiliation(s)}
\name{Hiroshi Seki$^{1}$ \qquad Takaaki Hori$^{2}$ \qquad Shinji Watanabe$^{3}$}
\address{$^{1}$ Toyohashi University of Technology, Japan. \\
$^{2}$ Mitsubishi Electric Research Laboratories (MERL), USA. \\
$^{3}$ Johns Hopkins University, USA.
}

\begin{document}
\ninept
\maketitle
\begin{abstract}

Attention-based encoder decoder network uses a left-to-right beam search algorithm in the inference step.
The current beam search expands hypotheses and traverses the expanded hypotheses at the next time step. This traversal is implemented using a for-loop program in general, and it leads to speed down of the recognition process.
In this paper, we propose a parallelism technique for beam search, which accelerates the search process by vectorizing multiple hypotheses to eliminate the for-loop program. 
We also propose a technique to batch multiple speech utterances for off-line recognition use, which reduces the for-loop program with regard to the traverse of multiple utterances.
This extension is not trivial during beam search unlike during training due to several pruning and thresholding techniques for efficient decoding.
In addition, our method can combine scores of external modules, RNNLM and CTC, in a batch as shallow fusion.
We achieved 3.7$\times$ speedup compared with the original beam search algorithm by vectoring hypotheses, and achieved 10.5$\times$ speedup by further changing processing unit to GPU.

\end{abstract}

\section{Introduction}
\label{sec:intro}

There is a great interest in automatic speech recognition (ASR) system because of the success of deep learning~\cite{xiong2017microsoft, audhkhasi2018building, chiu2018state, li2018advancing, hori2017interspeech} and popularization of speech interfaces, e.g., smart-phone and smart-speaker.
In practice, rapid execution of ASR decoding is essential for better user experience.
Reduction of sequence length~\cite{chan2016listen, hori2017interspeech, zeyer2018improved} and parallel computing~\cite{dixon2009harnessing, chong2009fully, chen2018gpu} are mainly investigated for rapid computation of likelihoods and efficient traversal of search space.

Beam search~\cite{aubert2002overview} is one of the breadth-first search algorithms which imposes a restriction on the search space to reduce the computational complexity of both memory space and execution time. During the search, hypotheses are expanded from a root node, and the expanded nodes at each depth level (or time-step in time-synchronous beam search for ASR) are stored in a FIFO (First-In First-Out) queue for further expansion at next depth level.
Thread parallelism and GPU-based execution accelerate computation of matrix multiplication and element-wise operation. However, the loop program with regard to hypothesis traversal still exists and decoder network needs to be executed per hypothesis in case of attention-based encoder decoder network. Therefore, there is a room for improvement of recognition time by concatenating hypotheses and processing them in a batch.

Dixon, et. al., proposed GPU based computation of acoustic scores~\cite{dixon2009harnessing}, and
Chong, et. al.,~\cite{chong2009fully} and Chen, et. al.,~\cite{chen2018gpu} further extended the search algorithm by executing graph traversal on GPU.
These studies focused on efficient computation of WFST (Weighted Finite-State Transducer) based decoding.

Different from the earlier works, 
we focus on a faster beam search algorithm for end-to-end attention-based encoder decoder networks.
We first vectorize $B$ beam size hypotheses and compute posterior probabilities for hypothesis expansion at next time step in a batch. 
This enables elimination of for-loop program with regard to beam size originally managed by FIFO (First-In First-Out) queue.
Next, we vectorize multiple $S$ input speech utterances to reduce the execution of for-loop program with regard to input speech data size.
It is not trivial unlike during training due to introduction of several pruning and thresholding techniques per utterance for efficient decoding.
During beam search, the encoder network generates hidden vectors of $S$ utterances at once, and the attention network and the decoder network process $S \times B$ hypotheses in a batch.
This algorithm is executable on both CPU and GPU without needing significant code modification.
In the experiment, we evaluate the effectiveness of the hypothesis and speech vectorization method assuming the following two scenarios:
\begin{itemize}
    \setlength{\leftskip}{-0.6cm}
    \item Online decoding\footnote{It is not a pure time-synchronous beam search because we used an attention-mechanism and bidirectional LSTM in the experiment. The proposed search algorithm is applicable to other online neural network architectures for pure time-synchronous beam search.}: $B$ hypotheses are vectorized to eliminate the loop for hypothesis traversal. Vectorization of hypotheses enables execution of attention and decoder networks for $B$ hypotheses in a batch.
    \item Offline decoding: $S$ speech utterances are further vectorized by processing multiple speech utterances. Vectorization of utterances and hypotheses enables execution of encoder network for $S$ utterances in a batch and also enables execution of attention and decoder networks for $S \times B$ hypotheses in a batch.
\end{itemize}

The rest of the paper is organized as follows.
Original implementation of beam search is described in Section~\ref{sec:beam}. Our contributions, vectorization of hypotheses and utterances, are described in Section~\ref{sec:batchbeam}. Experiments are conducted using librispeech corpus and CSJ corpus in Section~\ref{sec:exp}, followed by conclusions in Section~\ref{sec:conc}.

\section{Beam search}
\label{sec:beam}
\subsection{Definition}
Let $\mathcal{H}^{t} = (h^t_1, h^t_2, \cdots, h^t_b, \cdots h^t_B)$ be a set of hypotheses in the FIFO-queue at decoding time step $t$. Hypothesis $h^t_b$ has its own label history accumulated up to time step $t$:
\begin{align}
    h^t_b = l^1_{b} \cdot l^2_{b} \cdots l^t_{b},
\end{align}
where $l_b^k \in L$ denotes the $k$-th output label of $h_b^t$ in distinct output label set $L$.

At next time step $t+1$, the decoder network generates $|L|$ new labels with its posterior probabilities which leads to $|L| \times B$ hypotheses.
Let $\mathcal{L}=\{ i \in \mathbb{N} : i \le |L| \}$ be a set of indices for output labels, and $\mathcal{B}=\{ b \in \mathbb{N} : b \le B \}$ for current hypotheses.
Then the hypotheses at next time step $t+1$ are stored in a queue as the following equation:
\begin{align}
    \mathcal{H}^{t+1} \leftarrow \{ h^{t+1}_{b|L|+ i} | i \in \mathcal{L}, b \in \mathcal{B} \} \\
    \mathrm{where} \ \ h^{t+1}_{b|L|+ i} = h^{t}_{b} \cdot l^{t+1}_i
\end{align}
Each hypothesis has a score which is an accumulation of log posterior probability $\alpha$ up to decoding time step $t$, and it is updated by adding the output of decoder network:
\begin{align}
    \mathcal{Q}^{t+1} \leftarrow \{ &\alpha (h^{t+1}_{i|L|+b}) | i \in \mathcal{L}, b \in \mathcal{B} \} \\
    &\alpha (h^{t+1}_{i|L|+b}) = \alpha (h^{t}_{b}) + \log (p^{\mathrm{att}} (l_i^{t+1}))
\end{align}
where $p^{\mathrm{att}} (l^{t+1}_i)$ is the probability of label $l_i$ calculated by output of the decoder network.
Let $p^{\mathrm{att}}({l_{*}})$ be a set of posterior probabilities generated by the decoder network and $p^{\mathrm{att}}(l_i)$ be the posterior probability of $i$-th label.
In this paper, we follow the notation in~\cite{kim2016joint_icassp2017}:
\begin{align}
    p^{\mathrm{att}}(l_*^{t+1}) &= \mathrm{Generate}(c^{t+1}_b, r^{t}_{b}), \label{eq:dec1} \\
    r^{t+1}_{b|L|+i} &= \mathrm{Recurrency}(r^{t}_b, c^{t+1}_b, l^{t+1}_{i}), \label{eq:rec1}
\end{align}
where $r$ is the decoder state and $c$ is the context vector. Please refer to~\cite{kim2016joint_icassp2017} for detail.

For the reduction of search space, the expanded hypotheses are pruned at each time step.
In the experiment, we pruned the hypotheses in two step procedure, local pruning and global pruning.
At the local pruning, the log probabilities computed by the decoder network at time step $t$ are sorted in descending order, and top $B$ probabilities are selected as candidates.
When we define the function to select top $B$-candidates with its indices from the set of hypotheses $Q$ as $\mathrm{Select}(Q, B)$, the local pruning is represented as:
\begin{align}
    Q_b^{t+1'}, \ \zeta_b^{t+1'} = \mathrm{Select}\Bigl(\{\alpha (h^t_b) + \log p^{\mathrm{att}}(l^{t+1}_i) \ \mathrm{for} \ i \in \mathcal{L} \}, B\Bigr), \nonumber \\
    Q^{t+1'} = \{ Q_b^{t+1'} \ \mathrm{for} \ b \in \mathcal{B} \}, \ 
    \zeta^{t+1'} = \{ \zeta_b^{t+1'} \ \mathrm{for} \ b \in \mathcal{B} \}, \label{eq:loc_prune1}
\end{align}
resulting $B \times B$ hypotheses (and corresponding accumulated scores), where $\zeta$ is a set of selected indices. At the global pruning, they are further pruned to $B$ hypotheses as:
\begin{align}
    Q^{t+1''}, \zeta^{t+1''} = \mathrm{Select}\Bigl(Q^{t+1'}, B \Bigr).
\end{align}
Other search parameters, e.g., labels and cells in recurrent connections, are pruned for next time step by tracking the indices. When we define this function as $\mathrm{IndexSelect}$, the hypotheses, for example, is represented as:
\begin{align}
    \mathcal{H}^{t+1''} = \mathrm{IndexSelect}(\mathcal{H}^{t+1}, \zeta^{t+1'}, \zeta^{t+1''}). \label{eq:idxtrack}
\end{align}

\subsection{Implementation}
\label{sec:beam_impl}
The decoder network in Eq.~(\ref{eq:dec1}) takes previous label information at time step $t$ to output the posterior probabilities at time step $t+1$.
Other than the previous label, the networks with recurrent connection have its internal states (e.g., $r_b^t$ in Eq.~(\ref{eq:dec1}), $c_b^t$ in Eq.~(\ref{eq:rec1}), and attention weight) which will be used in a future time step.
These states also need to be pruned same as hypotheses.
At implementation level, each hypothesis is represented as a dictionary data structure consists of these states,
and stored in the FIFO-queue to reduce the execution of Eq.~(\ref{eq:idxtrack}).

\section{Hypotheses and speech vectorization}
\label{sec:batchbeam}
\subsection{Definition}

\begin{figure}
    \centering
    \includegraphics[width=7cm]{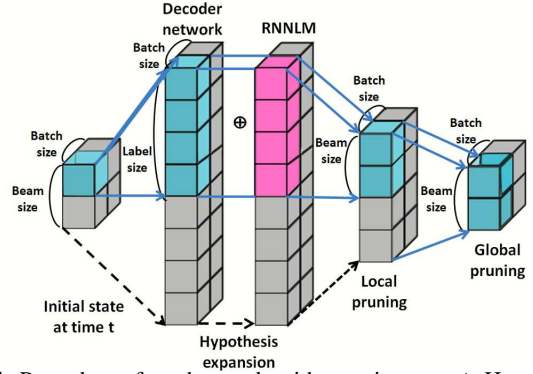}
    \vspace{-0.5cm}
    \caption{Procedure of our beam algorithm at time step $t$. Hypotheses are expanded and added with scores of RNNLM as shallow fusion. The candidate hypotheses are pruned by applying local and global pruning.}
    \label{fig:hyp_expantion}
\end{figure}

In this section, we reformulate the beam search algorithm in Section~\ref{sec:beam} by vectorizing the hypotheses and eliminating the loop with regard to beam size $B$. We further batch $S$ multiple utterances for the reduction of computational time assuming offline decoding scenario. In case of online decoding scenario, the batch size $S$ is set to 1.
Figure~\ref{fig:hyp_expantion} shows an overview of the proposed hypotheses expansion and pruning techniques at time step $t$.

For this purpose, we vectorize each element in the dictionary consists of the internal states as described in Section~\ref{sec:beam_impl}.
At time step $t=0$, the previous labels are defined as a vector of "start-of-sequence" symbols:
\begin{align}
    \mathbf{l}^0_{[S \times B]} = [\texttt{<sos>}, \cdots, \texttt{<sos>}]^\intercal,
\end{align}
and the accumulated scores are defined as:
\begin{align}
    Q^0_{[S \times B]} = [0.0, \cdots, 0.0]^\intercal,
\end{align}
The size of vector is represented in the square brackets.
By concatenating $S$ utterances, the encoder network can compute the hidden representations for $S$ utterances at once.
The output of encoder network is then duplicated to $B$ hypotheses to match the number of hypotheses.
Then, the decoder network computes the posterior probabilities for all $B$ beam hypotheses of $S$ utterances in a batch.
Let $\mathbf{\gamma}^t_{[S \times B, |L|]}$ be the calculated posterior probabilities for $S \times B$ candidates.
The attention-based decoder network in Eqs.~(\ref{eq:dec1}) and (\ref{eq:rec1}) are replaced as:
\begin{align}
    \mathbf{\gamma}^{t+1}_{[S \times B, |L|]} &= \mathrm{Generate}(\mathbf{c}^{t+1}_{[S \times B]}, \mathbf{r}^{t}_{[S \times B]}), \label{eq:dec2} \\
    \mathbf{r}^{t+1}_{[S \times B]} &= \mathrm{Recurrency}(\mathbf{r}^{t}_{[S \times B]}, \mathbf{c}^{t+1}_{[S \times B]}, \mathbf{l}^{t+1}_{[S \times B]}), \label{eq:rec2}
\end{align}

After the expansion of hypotheses, the local pruning is applied to reduce the number of hypotheses from $|L|$ to $B$ for all $B$ hypotheses and $S$ utterances. We define this function as $\mathrm{Select}(\mathbf{\gamma}^t_{[S \times B, |L|]}, B, \mathrm{idx})$ where $B$ is a number for return of top-$B$ candidates, and $\mathrm{idx}$ is a target index of selection\footnote{PyTorch supports this function as torch.topk.}. The selected log probabilities are added to the accumulated score. To match the dimension of the log probabilities and the accumulated score, we duplicated the accumulated score up to $B$ by introducing a new axis:
\begin{align}
    \mathbf{\alpha}^{t}_{[S \times B, B]} &\leftarrow \mathrm{Duplicate} (\mathbf{\alpha}^{t}_{[S \times B]}, B).
\end{align}
The accumulated score at time step $t+1$ is:
\begin{align}
    Q^{t+1}_{[S \times B, B]}, \ \mathbf{\zeta}'^{t+1}_{[S \times B, B]} &=
    \mathrm{Select}(\mathbf{\gamma}^{t+1}_{[S \times B, |L|]}, B, 2), \label{eq:loc_prune2} \\
    Q'^{t+1}_{[S \times B, B]} &= \mathbf{\alpha}^{t}_{[S \times B, B]} + Q^{t+1}_{[S \times B, B]}
\end{align}
where $\zeta'$ is the indices of top-$B$ output label candidates of $B$ hypotheses of $S$ utterances.
The accumulated score $Q'^{t+1}_{[S \times B, B]}$ is re-sized to $Q'^{t+1}_{[S, B \times B]}$ for global pruning targeting $B \times B$ candidates for $S$ utterances. 
The global pruning is represented as:
\begin{align}
    Q'^{t+1}_{[S, B \times B]} &= \mathrm{Resize}(Q'^{t+1}_{[S \times B, B]}), \\ 
    Q''^{t+1}_{[S, B]}, \mathrm{\zeta}''^{t+1}_{[S, B]} &= \mathrm{Select}(Q'^{t+1}_{[S, B \times B]}, B, 2).
\end{align}

Same as Section~\ref{sec:beam}, the other variables are pruned by tracking the selected indices.
In case of the hypotheses, the labels are duplicated to match the size of selected indices in Eq.~(\ref{eq:loc_prune2}): 
\begin{align}
    \mathbf{I}_{[S \times B, |L|]} &= \mathrm{Duplicate}(L, S\times B)^\intercal.
\end{align}
The duplicated labels are pruned and concatenated to update the hypotheses:
\begin{align}
    H^{t+1}_{[S \times B]} &= H^{t}_{[S \times B]} \nonumber \\ &\oplus \mathrm{IndexSelect}(\mathbf{I}_{[S \times B, |L|]}, \mathbf{\zeta}'^{t+1}_{[S \times B,B]}, \mathbf{\zeta}''^{t+1}_{[S,B]}),
\end{align}
where $\oplus$ is the operation for element-wise concatenation of accumulated label history and the current label. 

\subsection{Shallow fusion of external modules}
During beam search, scores of RNNLM (recurrent neural network language model) and CTC prefix score are integrated as shallow fusion. ESPnet~\cite{watanabe2018espnet} combines these scores and the final log probability, $\log p^{\mathrm{hyb}}$, is defined as weighted sum of CTC prefix score  ($p^{\mathrm{ctc}}$), decoder network ($p^{\mathrm{att}}$), and RNNLM ($p^{\mathrm{lm}}$):
\begin{align}
    \log p^{\mathrm{hyb}} (l_{t+1} | l_{1:t}, O)
    &= \lambda \log p^{\mathrm{ctc}} (l_{t+1} | l_{1:t}, O) \nonumber \\
    &+ (1 - \lambda) \log p^{\mathrm{att}} (l_{t+1} | l_{1:t}, O) \nonumber \\
    &+ \kappa \log p^{\mathrm{lm}} (l_{t+1} | l_{1:t}), \label{eq:p_hyb}
\end{align}
where $\lambda$ and $\nu$ are hyper-parameters and these values control contribution of each score.
Please refer to~\cite{watanabe2018espnet} for further detailed explanation.
The Eq.~(\ref{eq:dec2}) is rewritten as Eq.~(\ref{eq:p_hyb}) to combine the scores of RNNLM and CTC.

\section{Experiments}
\label{sec:exp}
\subsection{Experimental setup}

We used English and Japanese speech corpora, Librispeech~\cite{vassil2015librispeech} and CSJ (Corpus of Spontaneous Japanese)~\cite{maekawa2000spontaneous, maekawa2003corpus}\footnote{Recipes are available at ESPnet~\cite{watanabe2018espnet}.}.
As input feature, we used 80-dimensional log Mel filterbank coefficients 
and pitch features with its delta and delta delta features (80+3=83dimension)
extracted using Kaldi tools~\cite{Povey_ASRU2011}.
Joint CTC/attention-based encoder decoder networks~\cite{hori2017interspeech} were trained by using PyTorch~\cite{paszke2017automatic}.

On Librispeech corpus, we used a 8 layer BLSTM as the encoder network. The 2nd and 3rd bottom layers of the encoder network subsample hidden vector by the factor of 2~\cite{bahdanau2016end}.
Each BLSTM layer has 320 cells in each direction, and is followed by a linear projection layer with 320 units to combine the forward and backward LSTM outputs. The decoder network has an 1-layer LSTM with 300 cells. The number of labels was set to 29 including alphabets and special tokens.
On CSJ corpus, we used a 4 layer BLSTM as the encoder network with the subsampling technique. Each BLSTM  layer has 1024 cells in each direction, and is followed by a linear projection layer with 320 units to combine the forward and backward  LSTM  outputs. The decoder network has an 1-layer LSTM with 1024 cells. The number of labels was set to 3,260 including Japanese Kanji/Hiragana/Katakana characters and special tokens.

Beam search was performed using Intel Xeon Processor E5-2667 v3 for CPU-based search and Tesla K80 for GPU-based search. As evaluation set, we used randomly selected 1,000 utterances ($\approx 124.7$ minutes) on librispeech corpus and evalution set-1 ($\approx 110.1$ minutes) on CSJ corpus.

\subsubsection{Search parameters}
In the case of shallow fusion, we used $\lambda = 0.3$ and $\kappa = 0.3$ on both librispeech and CSJ. The beam size $B$ was set to 20 in decoding under all conditions.
For the recognition without vectorization, we conducted thread parallelism and process parallelism to accelerate decoding time.
In the case of thread parallelism, we controlled an environment variable $\mathrm{OMP\_NUM\_THREADS}$ and activated OpenMP.
We did not change other parameters and left it to the back-end PyTorch. In case of process parallelism, test data is split into multiple subsets and each subset is recognized in parallel using multiple CPU cores independently.

\subsection{Online decoding scenario}
Table~\ref{tab:online_libri} shows a duration (minutes) and real-time factor on Librispeech which parallelize hypotheses assuming online scenario. The row "$\mathrm{CPU^{conv}}$" shows durations of recognition time based on conventional beam search algorithm using attention decoder (ATT) and attention decoder with RNNLM (+RNNLM).
"batch" is a number of utterances $S$ for concatenation.
"threads" is a number of threads for thread parallelism and "procs" is a number of CPU cores used for process parallelism.
In the case of ATT, recognition time of original beam search was 318.5 minutes, and it was decreased to 85.0 minutes by parallelizing 20 beam hypotheses on CPU. It was further decreased to 30.4 minutes by changing processing unit to GPU. Recognition based on ATT+RNNLM also showed speed improvement.
Table~\ref{tab:online_csj} shows a duration (minutes) and real-time factor on CSJ corpus. The result on CSJ also showed the effectiveness of hypotheses vectorization with the usage of GPU for all conditions, ATT and ATT+RNNLM.

\begin{table}[tb]
    \caption{Duration (minutes) and real time factor in parenthesis on librispeech in online scenario}
    \label{tab:online_libri}
    \centering
    \scalebox{0.9}{
    \begin{tabular}{c|ccc|cc}
     & batch & threads&procs & ATT & +RNNLM \\ 
     \hline
     $\mathrm{CPU^{conv}}$ & 1 & 1&1 & \shortstack{318.5 \\ (2.6)} & \shortstack{518.3 \\ (4.2)} \\
     \hline
     CPU & 1 & 1&1 & \shortstack{85.0 \\ (0.7)} & \shortstack{108.2 \\ (0.9)} \\
     GPU & 1 & 1&1 & \shortstack{30.4 \\ (0.2)} & \shortstack{33.0 \\ (0.3)} \\
     \hline
    \end{tabular}
    }
\end{table}

\begin{table}[tb]
    \caption{Duration (minutes) and real time factor in parenthesis on CSJ in online scenario}
    \label{tab:online_csj}
    \centering
    \scalebox{0.9}{
    \begin{tabular}{c|ccc|cc}
     & batch & threads&procs & ATT & +RNNLM \\ 
     \hline
     ${\mathrm{CPU^{conv}}}$ & 1 & 1&1 & \shortstack{591.3 \\ (5.4)} & \shortstack{713.7 \\ (6.5)} \\ 
     \hline
     CPU & 1 & 1&1 & \shortstack{163.6 \\ (1.5)} & \shortstack{190.4 \\ (1.7)} \\
     GPU & 1 & 1&1 & \shortstack{32.2 \\ (0.3)} & \shortstack{32.2 \\ (0.3)} \\
     \hline
    \end{tabular}
    }
\end{table}

Our algorithm achieved significant gain from the conventional beam search algorithm on both librispeech corpus and CSJ corpus by vectorizing 20 hypotheses and eliminating the for-loop program for hypothesis traversal. In the case of ATT and ATT+RNNLM, real time factors were less than 1.0 and are applicable to online decoding scenario.

\subsection{Offline decoding scenario}

Table~\ref{tab:offline_libri} shows recognition time of thread parallelism (threads $>$ 1), process parallelism (procs $>$ 1), and our hypothesis and speech vectorization method (batch $>$ 1) on librispeech corpus. When we used 8 threads and decoded using decoder network, recognition time was comparable to the single thread execution as in Table~\ref{tab:online_libri}.

When multiple utterances are vectorized and recognized on CPU using the decoder network, the recognition time was 96.1 minutes. It was comparable to the process parallelism (80.3 minutes) even though our program consumed only one CPU core. The recognition time was further decreased to 16.0 minutes by changing the processing unit to GPU.
Comparison with Table~\ref{tab:online_libri} showed the advantage of utterance vectorization: 
in the case of GPU-based execution, recognition time without utterance vectorization was 30.4 minutes, however, vectorization of multiple utterances decreased the recognition time to 16.0 minutes.
In case of ATT+RNNLM, execution on one CPU core with vectorization of utterance and hypothesis consumed 104.8 minutes and it was comparable to the recognition time of process parallelism. 
Again, execution on GPU decreased the recognition time from 104.8 minutes to 16.1 minutes.

\begin{table}[tb]
    \caption{Duration (minutes) on librispeech in offline scenario}
    \label{tab:offline_libri}
    \centering
    \scalebox{0.9}{
    \begin{tabular}{c|ccc|cc}
     & batch & threads & procs & ATT & +RNNLM \\
    \hline
    \multirow{3}{*}{$\mathrm{CPU^{conv}}$} & 1 & 8 & 1 & 317.9 & 403.2 \\
    & 1 & 1 & 8 & 80.3 & 136.4  \\
    & 1 & 1 & 16 & 56.4 & 102.6 \\
    \hline
    \multirow{2}{*}{CPU} & 8 & 1&1 & 96.1 & 104.8 \\
     & 16 & 1&1 & 102.6 & 112.2 \\
    \multirow{2}{*}{GPU} & 8 & 1&1 & 16.0 & 16.1 \\
     & 16 & 1&1 & 15.2 & 14.5 \\
    \hline
    \end{tabular}
    }
\end{table}

\begin{table}[tb]
    \caption{Duration (minutes) on CSJ in offline scenario}
    \label{tab:offline_csj}
    \centering
    \scalebox{0.9}{
    \begin{tabular}{c|ccc|cc}
     & batch & threads & procs & ATT & +RNNLM \\
    \hline
    $\mathrm{CPU^{conv}}$ & 1 & 1 & 8 & 150.6 & 162.1 \\
    \hline
    CPU & 8 & 1 & 1 & 127.6 & 138.1 \\
    GPU & 8 & 1 & 1 & 16.1 & 17.0 \\
    \hline
    \end{tabular}
    }
\end{table}

Table~\ref{tab:offline_csj} shows the recognition time on CSJ.
When the recognition was performed using the score of decoder network, the recognition time was decreased from 591.3 minutes (in Table~\ref{tab:online_csj}) to 127.6 minutes,
and it was further decreased to 16.1 minutes by changing processing unit to GPU.

By vectorizing 8 multiple utterances, recognition time of our algorithm showed comparable performance with process parallelism with 8 CPU cores on both two corpora. In addition, execution based on GPU can fully exploit the advantage of GPU, and achieved further reduction of recognition time in case of ATT and ATT+RNNLM.

\subsection{Fusion of CTC prefix score}
\begin{table}[tb]
    \caption{Duration (minutes) on librispeech with shallow fusion of RNNLM and CTC prefix score.}
    \centering
    \label{tab:libri_ctc}
    \scalebox{0.9}{
    \begin{tabular}{c|ccc|c}
     & batch & threads & procs & ATT+RNNLM/CTC  \\
    \hline
    $\mathrm{CPU^{conv}}$ & 1 & 1 & 1 & 742.9 \\
    CPU & 1 & 1 & 1 & 205.0 \\
    GPU & 1 & 1 & 1 & 270.6 \\
    \hline
    $\mathrm{CPU^{conv}}$ & 1 & 1 & 8 & 134.3 \\
    CPU & 8 & 1 & 1 & 162.9 \\
    GPU & 8 & 1 & 1 & 51.3 \\
    \hline
    \end{tabular}
    }
\end{table}
Table~\ref{tab:libri_ctc} shows recognition time which use scores of RNNLM and CTC prefix score as shallow fusion. Recognition time of the original beam search was 742.9 minutes, and it was decreased by vectorizing hypotheses. Usage of GPU further decreased the recognition time to 270.6 minutes and achieved $2.7\times$ speedup.
We further vectorized 8 utterances in a batch. The recognition time was 51.3 minutes and it showed better result than the usage of 8 core CPU.

In the case of ATT+RNNLM/CTC, computation of CTC prefix score requires operations proportional to a length of hidden vector generated by the encoder network.
The $logsumexp$ operation in this computation slow down the speed especially when a large set of labels are used, and it was significant at CSJ corpus (3260 vs 29).
The recognition time of our algorithm based on GPU with 8-batch was 343.0 minutes and showed better result than the original program with single core CPU (742.9). However, it was slightly slower than the one with 8 core CPU (210.4). 
Acceleration of CTC prefix score is one of our future direction.

\section{Conclusions}
\label{sec:conc}
In this paper, we proposed a novel approach to speed up recognition time of beam search algorithm
by vectorizing search hypotheses and multiple input utterances.
We achieved 3.7$\times$ speedup compared with the original beam search algorithm by vectoring hypotheses on librispeech corpus, and 3.6 $\times$ speed up on CSJ corpus.
We further proposed a technique to batch multiple utterances.
In the case of GPU-based execution, 
vectorization of multiple utterances further achieved
1.9 $\times$ speed up on librispeech corpus and 2.0 $\times$ speed up on CSJ corpus.
This is available at open source project ESPnet.

\section{Acknowledgement}
We would like to thank Dr. Rohit Prabhavalkar at Google for many insightful discussions.

\bibliographystyle{IEEEbib}
\bibliography{strings,refs}

\begin{thebibliography}{10}

\bibitem{xiong2017microsoft}
Wayne Xiong, Jasha Droppo, Xuedong Huang, Frank Seide, Mike Seltzer, Andreas
  Stolcke, Dong Yu, and Geoffrey Zweig,
\newblock ``The {M}icrosoft 2016 conversational speech recognition system,''
\newblock in {\em IEEE International Conference on Acoustics, Speech and Signal
  Processing (ICASSP)}, 2017, pp. 5934--5938.

\bibitem{audhkhasi2018building}
Kartik Audhkhasi, Brian Kingsbury, Bhuvana Ramabhadran, George Saon, and
  Michael Picheny,
\newblock ``Building competitive direct acoustics-to-word models for english
  conversational speech recognition,''
\newblock in {\em IEEE International Conference on Acoustics, Speech and Signal
  Processing (ICASSP)}, 2018, pp. 4759--4763.

\bibitem{chiu2018state}
Chung-Cheng Chiu, Tara~N. Sainath, Yonghui Wu, Rohit Prabhavalkar, Patrick
  Nguyen, Zhifeng Chen, Anjuli Kannan, Ron~J. Weiss, Kanishka Rao, Ekaterina
  Gonina, Navdeep Jaitly, Bo~Li, Jan Chorowski, and Michiel Bacchiani,
\newblock ``State-of-the-art speech recognition with sequence-to-sequence
  models,''
\newblock in {\em IEEE International Conference on Acoustics, Speech and Signal
  Processing (ICASSP)}, 2018, pp. 4774--4778.

\bibitem{li2018advancing}
Jinyu Li, Guoli Ye, Amit Das, Rui Zhao, and Yifan Gong,
\newblock ``Advancing acoustic-to-word {CTC} model,''
\newblock in {\em IEEE International Conference on Acoustics, Speech and Signal
  Processing (ICASSP)}, 2018, pp. 5794--5798.

\bibitem{hori2017interspeech}
Takaaki Hori, Shinji Watanabe, Yu~Zhang, and Chan William,
\newblock ``Advances in joint {CTC}-{A}ttention based end-to-end speech
  recognition with a deep {CNN} encoder and {RNN}-{LM},''
\newblock in {\em Interspeech}, 2017, pp. 949--953.

\bibitem{chan2016listen}
Willian Chan, Navdeep Jaitly, Quoc Le, and Oriol Vinyals,
\newblock ``Listen, attend and spell: A neural network for large vocabulary
  conversational speech recognition,''
\newblock in {\em IEEE International Conference on Acoustics, Speech and Signal
  Processing (ICASSP)}, 2016, pp. 4960--4964.

\bibitem{zeyer2018improved}
Albert Zeyer, Kazuki Irie, Ralf Schlüter, and Hermann Ney,
\newblock ``Improved training of end-to-end attention models for speech
  recognition,''
\newblock in {\em Proc. Interspeech}, 2018, pp. 7--11.

\bibitem{dixon2009harnessing}
Paul~R Dixon, Tasuku Oonishi, and Sadaoki Furui,
\newblock ``Harnessing graphics processors for the fast computation of acoustic
  likelihoods in speech recognition,''
\newblock {\em Computer Speech \& Language}, vol. 23, no. 4, pp. 510--526,
  2009.

\bibitem{chong2009fully}
Jike Chong, Ekaterina Gonina, Youngmin Yi, and Kurt Keutzer,
\newblock ``A fully data parallel {WFST}-based large vocabulary continuous
  speech recognition on a graphics processing unit,''
\newblock in {\em Proc. Interspeech 2009}, 2009, pp. 1183--1186.

\bibitem{chen2018gpu}
Zhehuai Chen, Justin Luitjens, Hainan Xu, Yiming Wang, Daniel Povey, and
  Sanjeev Khudanpur,
\newblock ``A {GPU}-based {WFST} decoder with exact lattice generation,''
\newblock {\em arXiv preprint arXiv:1804.03243}, 2018.

\bibitem{aubert2002overview}
Xavier~L Aubert,
\newblock ``An overview of decoding techniques for large vocabulary continuous
  speech recognition,''
\newblock {\em Computer Speech \& Language}, vol. 16, no. 1, pp. 89--114, 2002.

\bibitem{kim2016joint_icassp2017}
Suyoun Kim, Takaaki Hori, and Shinji Watanabe,
\newblock ``Joint {CTC}-attention based end-to-end speech recognition using
  multi-task learning,''
\newblock in {\em IEEE International Conference on Acoustics, Speech and Signal
  Processing (ICASSP)}, 2017, pp. 4835--4839.

\bibitem{watanabe2018espnet}
Shinji Watanabe, Takaaki Hori, Shigeki Karita, Tomoki Hayashi, Jiro Nishitoba,
  Yuya Unno, Nelson Enrique~Yolta Soplin, Jahn Heymann, Matthew Wiesner, Nanxin
  Chen, Adithya Renduchintala, and Tsubasa Ochiai,
\newblock ``{ESPnet}: end-to-end speech processing toolkit,''
\newblock {\em arXiv preprint arXiv:1804.00015}, 2018.

\bibitem{vassil2015librispeech}
Vassil Panayotov, Guoguo Chen, Daniel Povey, and Sanjeev Khudanpur,
\newblock ``{LIBRISPEECH}: An {ASR} corpus based on public domain audio
  books,''
\newblock in {\em IEEE International Conference on Acoustics, Speech and Signal
  Processing (ICASSP)}, 2015, pp. 5206--5210.

\bibitem{maekawa2000spontaneous}
Kikuo Maekawa, Hanae Koiso, Sadaoki Furui, and Hitoshi Isahara,
\newblock ``Spontaneous speech corpus of {Japanese},''
\newblock in {\em International Conference on Language Resources and Evaluation
  (LREC)}, 2000, vol.~2, pp. 947--952.

\bibitem{maekawa2003corpus}
Kikuo Maekawa,
\newblock ``Corpus of {Spontaneous} {Japanese}: Its design and evaluation,''
\newblock in {\em ISCA \& IEEE Workshop on Spontaneous Speech Processing and
  Recognition}, 2003.

\bibitem{Povey_ASRU2011}
Daniel Povey, Arnab Ghoshal, Gilles Boulianne, Lukas Burget, Ondrej Glembek,
  Nagendra Goel, Mirko Hannemann, Petr Motlicek, Yanmin Qian, Petr Schwarz, Jan
  Silovsky, Georg Stemmer, and Karel Vesely,
\newblock ``The kaldi speech recognition toolkit,''
\newblock in {\em IEEE Workshop on Automatic Speech Recognition and
  Understanding (ASRU)}, Dec. 2011.

\bibitem{paszke2017automatic}
Adam Paszke, Sam Gross, Soumith Chintala, Gregory Chanan, Edward Yang, Zachary
  DeVito, Zeming Lin, Alban Desmaison, Luca Antiga, and Adam Lerer,
\newblock ``Automatic differentiation in {P}y{T}orch,''
\newblock in {\em NIPS-W}, 2017.

\bibitem{bahdanau2016end}
Dzmitry Bahdanau, Jan Chorowski, Dmitriy Serdyuk, Philemon Brakel, and Yoshua
  Bengio,
\newblock ``End-to-end attention-based large vocabulary speech recognition,''
\newblock in {\em IEEE International Conference on Acoustics, Speech and Signal
  Processing (ICASSP)}, 2016, pp. 4945--4949.

\end{thebibliography}

\end{document}